\documentclass[12pt,preprint]{aastex}

\begin{document}

\def\arcsecpoint{$''\!.$}
\def\arcminpoint{$'\!.$}
\def\deg{$^{\rm o}$}
\def\ltsim{\raisebox{-.5ex}{$\;\stackrel{<}{\sim}\;$}}
\def\gtsim{\raisebox{-.5ex}{$\;\stackrel{>}{\sim}\;$}}



\shortauthors{Crenshaw, Kraemer, \& Gabel}
\shorttitle{Host Galaxies of Narrow-Line Seyfert 1s}

\title{The Host Galaxies of Narrow-Line Seyfert 1s: Evidence for
Bar-Driven Fueling}

\author{D. M. Crenshaw\altaffilmark{1},
S. B. Kraemer\altaffilmark{2},
\& J.R. Gabel\altaffilmark{2}}

\altaffiltext{1}{Department of Physics and Astronomy, Georgia State University,
Atlanta, GA 30303; crenshaw@chara.gsu.edu}

\altaffiltext{2}{Catholic University of America,
and Laboratory for Astronomy and Solar Physics, NASA's Goddard Space Flight 
Center, Code 681,
Greenbelt, MD  20771; stiskraemer@yancey.gsfc.nasa.gov.}

\begin{abstract}

We present a study of the host-galaxy morphologies of narrow- and broad-line 
Seyfert 1 galaxies (NLS1s and BLS1s) based on broad-band optical images from 
the {\it Hubble Space Telescope} archives. We find that large-scale stellar 
bars, starting at $\sim$1 kpc from the nucleus, are much more common in NLS1s 
than BLS1s. Furthermore, the fraction of NLS1 spirals that have bars increases 
with decreasing full-width at half-maximum (FWHM) of the broad component of 
H$\beta$. These results suggest a link between the large-scale bars, which can 
support high fueling rates to the inner kpc, and the high mass-accretion rates 
associated with the supermassive black holes in NLS1s.

\end{abstract}


\section{Introduction}

Seyfert galaxies are relatively nearby ($z \lesssim$ 0.1), moderate-luminosity 
($L_{bol}$ $=$ 10$^{43}$ -- 10$^{45}$ erg s$^{-1}$) active galaxies. The 
large-scale morphologies of these galaxies typically resemble those of 
normal (i.e., inactive) spiral galaxies, although there are some examples 
of peculiar or interacting systems (De Robertis, Hayhoe, \& Yee 1998). Optical 
and UV spectra of their nuclei are characterized by strong atomic emission 
lines. Based on the widths of these lines, Seyfert galaxies are generally 
divided into two types (Khachikian \& Weedman 1974): Type 1s possess broad 
permitted lines with full-width at half-maximum (FWHM) typically $>$ 1000 km
s$^{-1}$ and narrow forbidden lines with FWHM $\ltsim$ 500 km s$^{-1}$, 
while the spectra of Type 2s show narrow permitted and forbidden lines. Those 
Seyferts that show both broad and narrow components in their permitted lines 
are classified as intermediate types, ranging from 1.2 to 1.9 (Osterbrock, 
1977; 1981). 

Osterbrock \& Pogge (1985) discovered a class of active galactic nuclei (AGN) 
with relatively narrow 
(FWHM $<$ 2000 km s$^{-1}$) permitted lines, like Seyfert 2 galaxies, but 
with emission-ratios that indicate the presence of a high-density region, like 
the broad-line region (BLR) in Seyfert 1 galaxies.
They dubbed these objects ``Narrow-Line Seyfert 1s'' (NLS1s).
As discussed by Mathur (2000a, 2000b), NLS1s possess a number of interesting
spectral properties. They show strong Fe~II and weak [O~III] $\lambda$5007 
emission relative to H$\beta$, which puts them at one extreme of eigenvector 1 
of Boroson \& Green (1992), which was determined from principal component 
analysis of a large sample of low-redshift quasars and Seyfert galaxies. 
Compared to broad-line Seyfert 1s (BLS1s), the X-ray continua of NLS1s are 
characterized by steep slopes and rapid variability, particularly in the soft 
($\leq$ 1 keV) band (Boller, Brandt, \& Fink 1996; Brandt, Mathur, \& Elvis 
1997). Wills et al. (1999) found unsually high N~V $\lambda$1240/ C~IV 
$\lambda$1550 flux ratios in NLS1s, which may be indicative of high nitrogen 
abundances, as would result from metal enrichment due to a recent burst of star 
formation (see also Shemmer \& Netzer 2002). The mid-IR brightness of a number 
of NLS1s provides further evidence of recent or current star formation
(Moran, Halpern, \& Helfand 1996). 

There are several models to explain the unusual properties of NLS1s, and in 
particular the narrowness of their line profiles. Wandel \& Peterson (1999) 
suggested that the BLRs may be relatively farther away from the central 
continuum source in NLS1s, due to 
over-ionization of the gas close to the active nucleus. However,
based on estimates of the BLR sizes from reverberation mapping, Peterson et al. 
(2000) argue this cannot generally be the case. Another possibility is that the 
BLR is flattened and viewed roughly face-on, with narrow line profiles as a 
result, although Peterson et al. found no compelling evidence for such a 
preferred viewing angle unless the narrow emission-line regions are mis-aligned 
with the accretion-disk/black hole systems in these objects.

The current widely-accepted paradigm is that, while AGN in general are powered 
by accretion of material onto a supermassive central black hole, NLS1s possess 
black holes of relatively modest mass ($\leq$ 10$^{7}$ M$_\odot$), that are 
accreting matter at or above their Eddington limits (Pounds, Done, \& Osborne 
1995). This view is supported by recent observational results that indicate that
NLS1s possess significantly smaller black-hole mass to bulge mass ratios 
than their broad-line counterparts (Mathur, Kuraszkiewicz, \& Czerny 2001; 
Wandel 2002). In this scenario, the narrow line widths are simply due to clouds 
in motion around the small-mass 
black hole, while the steep soft X-ray continuum is the high-energy tail of the 
``Big Blue Bump'', which is presumably emission from the accretion
disk that peaks in the extreme ultraviolet (EUV). NLS1s could therefore be 
analogous to the ``strong soft'' states of galactic black holes at high 
accretion rates (Pounds et al. 1995).
Due to the high accretion rates, the disks in NLS1s are likely much hotter than 
those in BLS1s, and hence the emission is peaked at higher energies. Wang \& 
Netzer (2003) showed that high-accretion rates result in an extremely thin disk 
which emits a double-peaked continuum, with a peak in the EUV/soft X-ray due to 
the disk itself, and a high energy peak due to a hot corona. Based on this 
model, they suggest that most of the 54 NLS1s in the V\'eron-Cetty, V\'eron, \& 
Gon\c{c}alves (2001) sample are super-Eddington accretors. Assuming that NLS1s 
are 
high-rate accretors, and considering the other evidence of extreme activity, 
Mathur (2000a, 2000b) suggested that NLS1s are in an early phase of activity 
and may be 
the low-redshift analogs of high-$z$ QSOs (however, see Constantin \& Shields 
2003).   

In order to fuel Seyfert galaxies and other AGN, material must be transported 
from the host galaxy into 
the inner nucleus, requiring some means by which the material loses 
angular momentum. One possibility is interactions or mergers with other nearby 
galaxies (Toomre \& Toomre 1972; Adams 1977). However recent studies indicate 
that Seyfert galaxies do not have more close companions than do normal 
galaxies, and, furthermore, a substantial fraction of Seyfert galaxies show no 
evidence for recent mergers (De Robertis, Yee, \& Hayhoe 1998, and references 
therein).

Another process for fueling AGN that has received considerable attention is gas 
inflow along a stellar bar (Simkin, Su, \& Schwarz 1980). Theoretical studies 
show that the gas can lose its angular momentum after encountering the 
gravitational potential of a bar and be transported inward (Shlosman, Begelman, 
\& Frank 1989). However, most observational studies in this area have found 
similar fractions 
of bars for Seyferts and normal galaxies (Heckman 1980; Simkin et al.
1980; Ho, Filippenko, \& Sargent 1997; Mulchaey, Regan, \& Kundu 1997; Mulchaey 
\& Regan 1997), although Knappen, Shlosman, \& Peletier (2000) found a 
slightly higher percentage of IR bars in Seyferts (80\%) compared to normal 
galaxies (60\%). Interestingly, the percentages of Seyfert or normal galaxies 
that have bars vary substantially among these studies, from $\sim$30\% to 
$\sim$80\%. These discrepancies are probably due to the sample selection (e.g., 
luminous vs. average spirals), wave band (IR observations tend to reveal more 
bars than optical images, Mulchaey \& Regan 1997), and identification criteria 
(e.g., whether or not to include weak bars). 
Nevertheless, well-defined samples that use consistent criteria 
nearly always find equal percentages of bars in Seyfert and normal galaxies 
(with the exception noted above).
Since there is strong evidence that most normal galaxies contain 
inactive supermassive black holes (Kormendy \& Richstone 1995), the above 
finding indicates that while bars may be important for driving gas into the 
inner regions, there are other factors that contribute to the presence of 
nuclear activity.

Dynamical models indicate that a bar potential can drive material to within 
$\sim$1 kpc from the AGN, at which point the transport will be halted at the 
Inner Lindblad Resonance and the infalling gas will form a disk (Shlosman, 
et al. 1989). If global nonaxisymmetrical
instabilities are present, a secondary, gaseous bar can develop at small 
distances from the nucleus ($\ltsim$1 kpc), and provide a 
means to drive the gas in towards the active nucleus (Shlosman, et al. 1989; 
Heller \& Shlosman 1994). In fact, Maiolino et al. (2000) found 
evidence for gas motion along a secondary bar in the nucleus of the Seyfert 2 
Circinus Galaxy. However, it appears that secondary gas bars are fairly rare 
among Seyfert galaxies. The most efficient way to detect these bars is 
through the extinction or reddening caused by their embedded dust.
{\it Hubble Space Telescope} ({\it HST}) IR and optical images indicate that 
only 10 -- 20\% of Seyfert 
galaxies show nuclear (at distances $\ltsim$1 kpc from the nucleus) dust bars 
(Regan \& Mulchaey 1999; Martini \& Pogge 1999; Pogge \& Martini 2002).
Instead, the {\it HST} images reveal that most ($\sim$80\%) Seyfert galaxies 
show nuclear dust spirals (Regan \& Mulchaey 1999; Pogge \& Martini 2002; 
Martini et al. 2003). Although these structures are common to both barred and 
unbarred Seyferts, all of the ``grand design'' nuclear spirals (so-called by 
their resemblance to the large scale spiral structure of the same name)
are found in barred galaxies (Martini et al. 2003). The grand design spirals 
appear to connect to dust lanes on the leading edge of the large-scale stellar 
bars, suggesting a connection to shock fronts in the bars. However, Martini et 
al. (2003) find that dust spirals in Seyfert galaxies are not statistically 
more numerous than those in normal galaxies. Thus, it is 
unclear exactly how the dust spirals fuel the AGN and what mechanism(s) 
actually control the onset of nuclear activity.
  
If the high accretion-rate paradigm for NLS1s is correct, it suggests 
that the fueling of the AGN is more efficient in these Seyferts than in their 
BLS1 counterparts. If so, perhaps there is a connection between 
large-scale properties and the high accretion rates in NLS1s. Krongold, 
Dultzin-Hacyan, and Marziani (2001) examined host 
galaxies and environments, and found no statistical difference between the 
frequency of companion galaxies in NLS1s and BLS1s. Another possibility is a 
difference in their large-scale morphologies. In this paper, we provide 
evidence that, stellar bars are much more common in NLS1s than BLS1s. These 
large-scale bars, which typically begin at $\sim$1 kpc from the nucleus and 
extend to 5 -- 10 kpc, 
represent an efficient means for transporting large amounts of gas to the inner 
regions, which can presumably support the high accretion rates in the nuclei 
of NLS1s.

\section{Sample and Analysis}

Our sample contains primarily the {\it HST} broad-band images of Seyfert 1 
galaxies obtained by Malkan, Gorjian, \& Tam (1998). This is a uniform 
sample of 91 Seyfert 1 galaxies at z $\leq$ 0.035 observed with the Wide Field 
Planetary Camera 2 (WFPC2) through the F606W filter (centered near 6000 \AA). 
Single exposures of duration 500 s were obtained with nearly all the Seyfert 
nuclei located on the planetary camera (PC) chip, yielding a spatial resolution 
of 0\arcsecpoint1. The number of NLS1s in this sample is small (13); to 
increase that number, we searched the {\it HST} archives for NLS1s in the 
V\'eron-Cetty et al. (2001) sample that have WFPC2 broad-band optical images.
We found 6 additional NLS1s, which extend to higher redshifts (up to z 
$=$ 0.084); the effects of including them to make a larger, more heterogeneous 
sample (in terms of redshift) will be discussed.

We retrieved all of the WFPC2 images from the {\it HST} archives, to display 
a larger field of view (37$''$ x 37$''$, the projected size of the PC) than the 
images published (mostly 9$''$ x 9$''$) in Malkan et al. (1998). The Malkan et 
al. sample consists of single images of each Seyfert galaxy, and we removed 
only point-like cosmic-ray hits using automated routines. The remaining 
galaxies had multiple images, and we removed both point-like and glancing hits 
through intercomparison of the images. In practice, residual or remaining 
cosmic ray hits, saturation of the nuclei, or other defects were easily 
recognized in these images and had little effect on our assessment of the 
large-scale morphology of the host galaxies.

To ensure objectivity, we scrambled the order of the {\it HST} images and the 
three coauthors independently classified their morphologies 
without knowledge of Seyfert 1 subclass (NLS1 or BLS1) or the name of the 
galaxy. We classified the galaxies into one of 6 major groups: S (spiral), 
SB (spiral with a noticeable bar), E (elliptical), I (irregular), P 
(point-souce, only the active nucleus is visible), or ? (uncertain) (note that 
we did not attempt to assign Hubble subclasses).
For each galaxy, at least two of the three independent classifications 
agreed (in most cases all three agreed), and we adopted the majority view.

A large majority of the galaxies in the sample are at redshifts z $=$ 0.01 -- 
0.035, which yield projected distances of 7 -- 25 kpc across the PC for 
H$_0$ $=$ 75 km s$^{-1}$ Mpc$^{-1}$. In the Malkan sample, there are 11 out of 
91 galaxies at smaller redshifts, extending down to z $=$ 0.002 (1.4 kpc across 
the PC). For these galaxies in particular, we also examined the full WFPC2 
images to assess their large-scale morphologies (each WF chip has a field of 
view of 1\arcminpoint3 x 1\arcminpoint3). The six additional galaxies from 
V\'eron-Cetty et al. (2001) have redshifts up to 0.084, corresponding to 60 
kpc across the PC.

In Table 1, we give the results of our classifications. The galaxies are first 
listed in the same order as in Malkan et al. (1998), and are followed by the 
six additional NLS1 galaxies from V\'eron-Cetty et al. (2001). We also list the 
redshifts, our morphological classifications (``CKG''), and those based on the 
same images from Malkan et al. (``MGT''), which include Hubble subclasses. The 
classes of Seyfert nuclei (BLS1 
or NLS1) are based on the listings in the catalog of V\'eron-Cetty \& V\'eron 
(2001); the BLS1s include Seyferts from types 1 to 1.9. In the last two 
columns, we give measurements of the full-width at 
half-maximum (FWHM) of the broad component of H$\beta$ (or in a few cases,
H$\alpha$), and the references that these measurements were obtained from. 
Note that 35 Seyfert galaxies in the sample have previously been classified as 
NLS1s or BLS1s on 
the basis of their optical spectra, but measurements of the FWHMs of the broad 
lines have not been published.

We note that our general morphological classes 
agree well with those of Malkan et al. Of the 91 galaxies, there are only 11 
discrepancies, and of the 74 galaxies that we both identified as spirals, there 
is disagreement on the presence or absence of a bar in only 6 galaxies (5 
BLS1s, 1 NLS1).
To illustrate our classifications, we show the {\it HST} images of the NLS1s in 
our sample in Figures 1 and 2. NGC~4051 is not shown, since at its very low 
redshift (z $=$ 0.0023) the PC only spans 1.6 x 1.6 kpc. Examination of the 
Palomar Sky Survey and other ground-based images (Tully et al. 1996) shows a 
strong bar that can be seen extending into the larger WFPC2 frame. 

\section{Results}

We consider the 92 Seyfert galaxies in Table 1 with definite morphological 
classifications (i.e., ignoring the ``P'' and ``?'' classes). From this group, 
we have classified 91\% as spirals, and of the 
spirals, we have classified 33\% of the spirals as barred. The latter is 
consistent with percentages from studies that have identified strong bars in 
optical images of both active and normal spiral galaxies (Mulchaey et al. 1997; 
Malkan et al. 1998; see also \S1).

However, it is obvious from Table 1 that most of the NLS1 galaxies are barred 
spirals, and most of the BLS1 galaxies are normal spirals. Quantitatively, if 
we consider the 84 spirals in this sample, then 65\% (11/17) of the NLS1 
spirals have bars, whereas only 25\% (17/67) of the BLS1 spirals have bars. 
If we consider only the Malkan et al. sample, then 64\% (7/11) of the NLS1 
spirals have bars, which is essentially identical to the result for the full 
sample. If we use Malkan et al.'s (1998) classifications, then 30\% (20/67) of 
the BLS1 spirals show bars and 55\% (6/11) of the NLS1 spirals show bars.

To further characterize the trend of bars in NLS1 galaxies, Figure 3 shows 
histograms of the fraction 
of spirals that have bars as a function of the FWHM (H$\beta$) on a log scale.
The shaded regions show a clear trend of decreasing fraction of bars with 
increasing FWHM for the entire sample. The trend continues through the three 
bins associated with NLS1s (FWHM $<$ 2000 km s$^{-1}$) to the first bin for the 
BLS1s. The fraction levels out thereafter, and is essentially constant for all 
BLS1s regardless of FHWM. Interestingly, the four NLS1 spirals in the full 
sample with the smallest FWHM (600 -- 1000 km s$^{-1}$) all show bars.

The hatched region in Figure 3 shows that the same trend occurs for the more 
uniform Malkan et al. subsample; there is only a slightly lower fraction of 
bars in NLS1s with widths in the range 1000 -- 2000 km s$^{-1}$ compared to the 
full sample. 
It is clear from Table 1 that the number of NLS1s in each bin is quite 
small (as low as 3) for this subsample, and a larger, more uniform sample of 
images for both BLS1s and NLS1s would be helpful for testing this trend.

\section{Discussion}

Our analysis of {\it HST}/WFPC2 images reveals that stellar bars are
more common in NLS1s, particularly those with very narrow H$\beta$ profiles, 
compared to BLS1s. 
This suggests a link between the high fueling rates that can be provided by the 
bars and the high accretion rates commonly associated with NLS1s.
These large-scale stellar bars often extend down to 1 kpc from the nuclei of 
these galaxies. Inside a radius of $\sim$1 kpc, nuclear dust spirals, and in a 
few cases dust bars, are likely to be directly 
responsible for transporting fuel to the active nucleus. 
However, the exact connection between large-scale bars, inner dust spirals and 
bars, and accretion rates in the nucleus is unclear. We outline one possible 
scenario below.

Based on a variety of evidence, it has been suggested that NLS1s are in an 
early stage of activity (Mathur 2000a, 2000b).
If so, the presence of bars suggests the following evolutionary sequence for 
galaxies that become active.
1) Initially, the galaxy is a normal spiral, albeit
with a small black hole mass M$_{BH}$ ($<$ 10$^{7}$ M$_\odot$). Although 
studies of nearby 
inactive galaxies find few with M$_{BH}$ this small (Magorrian et al. 1998; 
Merritt \& Ferrarese 
2001; Kormendy \& Gebhardt
2001), Wandel (2002) suggested that the sample derived from stellar dynamics is 
biased towards detecting
higher mass M$_{BH}$, since detecting smaller mass sources requires higher 
spatial resolution, which is
only possible in the most nearby galaxies. Hence there may be a large, 
undetected population of small
M$_{BH}$ in inactive galaxies. 2) Due to either a tidal disruption or simply an 
asymmetrical distribution of mass
in the inner disk of the galaxy, a stellar bar forms within a few x 10$^{8}$ 
yrs (Combes \& Elmegreen 1993).
The presence of a bar can efficiently drive gas
into the inner nucleus (Friedli \& Benz 1993),
which can be detected in the form of inner dust spirals or bars.
3) At this point, the galaxy starts the 
NLS1 phase, powered by a small central
black hole accreting at close to its Eddington limit. Within 10$^{8}$ yrs, the 
bar can trigger star formation (Hunt \&
Malkan 1999, and references therein), which suggests that the evidence for 
active star formation in NLS1s
may also be related to the presence of a bar. Dynamical models indicate that 
once $\sim$ 5\% of the mass of 
the disk is redistributed into the inner
nucleus, the bar will be destroyed (Freidli \& Benz 1993; Norman et al. 1996). 
Although models predict that this
will take a few x 10$^{9}$ yrs, they do not typically include the effects of 
asymmterical mass distribution in the inner nucleus,
which may shorten the lifetime of the bar (Norman et al. 1996).
4) At this stage, 
M$_{BH}$ will have increased sufficiently
(several to ten times its initial mass) such that the galaxy appears as a 
BLS1. Since BLS1s are substantially sub-Eddington,
the fueling rate must also have decreased significantly. There are a number of 
possible explanations for this, including
the weakening of the bar, the diminished reserve of fuel in the inner disk, or 
the effect of the energized AGN
itself, which can drive mass outflow (see Crenshaw, Kraemer, \& George 2003).
Interestingly, Hunt \& Malkan 
(1999) find evidence that Seyferts show outer stellar rings
more often than inactive galaxies. Outer rings take $\sim$ several x 10$^{9}$ 
yrs to form, and are likely to outlast the stellar
bars. Hence, the rings may be a remnant of the earlier, bar-driven phase of 
what are now BLS1s.

Testing the above scenario, and determining the physical connection between 
large-scale bars and fueling of active nuclei, could be accomplished with more 
extensive {\it HST} images of the nuclear regions in NLS1s. The fueling flow 
can be traced via deep optical and IR images of the dust lanes, and possibly
narrow-band images of the low-ionization emission. Narrow-band observations of 
the high-ionization outflowing gas could be used to determine the geometry of 
infall versus outflow. Finally, UV images could be used to look for evidence of 
circumnuclear starbursts and explore their connection to AGN activity.

\acknowledgments

This research has made use of the NASA/IPAC Extragalactic Database (NED) which 
is operated by the Jet Propulsion Laboratory, California Institute of 
Technology, under contract with the National Aeronautics and Space 
Administration. This research has also made use of NASA's Astrophysics Data 
System Abstract Service. Some of the data presented in this paper were obtained 
from the Multimission Archive at the Space Telescope Science Institute (MAST). 
Support for MAST for non-HST data is provided by the NASA Office of Space 
Science via grant NAG5-7584 and by other grants and contracts.

\clearpage

\figcaption[f1a.eps]{{\it HST} WFPC2 (PC) images of NLS1s in the Malkan et al. 
(1998) sample. The field of view has been slightly trimmed to 34$''$ x 34$''$. 
Next to the name of each galaxy, the projected distances across the field of 
view are given assuming H$_0$ = 75 km s$^{-1}$ Mpc$^{-1}$.}

\figcaption[f2.eps]{{\it HST} WFPC2 (PC) images of NLS1s from V\'eron-Cetty et 
al (2001). The field of view has been trimmed to 18\arcsecpoint4 x 
18\arcsecpoint4 to show the large-scale structure more clearly. Next to the 
name of each galaxy, the projected distances across the field of view are given 
assuming H$_0$ = 75 km s$^{-1}$ Mpc$^{-1}$.}

\figcaption[f3.eps]{Histogram showing the fraction of Seyfert spirals with bars 
as a 
function of the full-width at half-maximum (FWHM) of the broad component of the 
H$\beta$ emission line. The shaded region is for the full sample, whereas the 
hatched region is for the Malkan et al. (1998) subsample.}

\newpage
\begin{deluxetable}{lcllccc}
\tablecolumns{7}
\footnotesize
\tablecaption{Properties of Seyfert 1 Galaxies in Sample}
\tablewidth{0pt}
\tablehead{
\colhead{Name} &
\colhead{Redshift} &
\colhead{Morph.} &
\colhead{Morph.} &
\colhead{Class} &
\colhead{FWHM} &
\colhead{Ref.$^a$} \\
\colhead{} &
\colhead{} &
\colhead{CKG} &
\colhead{MGT} &
\colhead{} &
\colhead{(km s$^{-1}$}) &
\colhead{} 
}
\startdata
ESO215-G14       & 0.019 & S  & Sa      & BLS1 &   &   \\
ESO323-G77       & 0.015 & S  & Sa/b    & BLS1 & 2500 & W92 \\
ESO354-G4        & 0.033 & S  & Sa      & BLS1 &   &   \\
ESO362-G18       & 0.013 & S  & Sa      & BLS1 & 4000 & W92 \\
ESO438-G9        & 0.024 & SB & SBc/d   & BLS1 & 5000 & K83 \\
F51              & 0.014 & S  & Sa      & BLS1 & 2700 & W92 \\
F1146            & 0.032 & S  & Sb      & BLS1 & 4300 & W92 \\
HEAO-1-0307-730  & 0.028 & SB & S(B)a   & BLS1 & 2900 & W92 \\
HEAO1143-181     & 0.033 & I  & I       & BLS1 & 2400 & W92 \\
HEAO2106-098     & 0.027 & P  & P       & BLS1 & 3835 & R00 \\
IC1816           & 0.017 & SB & SBa/b   & BLS1 &   &   \\
IC4218           & 0.019 & S  & Sa      & BLS1 &   &   \\
IC4329A          & 0.016 & S  & Sa      & BLS1 & 4800 & W92 \\
IR1319-164       & 0.017 & S  & Sb      & BLS1 &   &   \\
IR1333-340       & 0.008 & S  & SO      & BLS1 & 2400 & W92 \\
MCG6-26-12       & 0.032 & SB & SB0     & NLS1 & 1145 & V01 \\
MCG8-11-11       & 0.020 & S  & SB0     & BLS1 & 3630 & O82 \\
MARK6            & 0.019 & S  & S0      & BLS1 &   &   \\
MARK10           & 0.030 & S  & Sa/b    & BLS1 & 2400 & O77 \\
MARK40           & 0.020 & ?  & S0      & BLS1 & 2000 & O77 \\
MARK42           & 0.024 & SB & SBa     & NLS1 &  865 & V01 \\
MARK50           & 0.023 & S  & S0      & BLS1 &   &   \\
MARK79           & 0.022 & SB & SBc     & BLS1 & 4950 & O82 \\
MARK279          & 0.031 & S  & Sa      & BLS1 & 6860 & O82 \\
MARK290          & 0.029 & S  & E       & BLS1 & 2550 & O82 \\
MARK334          & 0.022 & S  & I       & BLS1 &   &   \\
MARK335          & 0.025 & P  & ?       & NLS1 & 1350 & O82 \\
MARK352          & 0.015 & E  & E       & BLS1 & 3800 & O77 \\
MARK359          & 0.017 & SB & SBb/c   & NLS1 &  900 & V01 \\
MARK372          & 0.031 & S  & Sa      & BLS1 & 5500 & G91 \\
MARK382          & 0.034 & SB & SBa     & NLS1 & 1280 & V01 \\
MARK423          & 0.032 & S  & Sb      & BLS1 & 9000 & O81 \\
MARK471          & 0.034 & SB & SBc     & BLS1 &   &   \\
MARK493          & 0.031 & SB & S(B)a   & NLS1 &  740 & V01 \\
MARK516          & 0.028 & S  & Sc      & BLS1 & 4000 & O81 \\
MARK530          & 0.029 & S  & Sa      & BLS1 & 6560 & K00 \\
MARK543          & 0.026 & S  & Sc      & BLS1 &   &   \\
MARK590          & 0.027 & S  & Sa      & BLS1 & 2680 & S90 \\
MARK595          & 0.028 & S  & Sa      & BLS1 & 2360 & S90 \\
MARK609          & 0.032 & S  & Sa/b    & BLS1 &   &   \\
MARK699          & 0.034 & E  & E       & NLS1 &  840 & O81 \\
MARK704          & 0.029 & SB & SBa     & BLS1 & 5500 & S90 \\
MARK744          & 0.010 & S  & Sb      & BLS1 &   &   \\
MARK766          & 0.012 & SB & SBc     & NLS1 & 1630 & V01 \\
MARK817          & 0.033 & SB & SBc     & BLS1 & 4300 & O82 \\
MARK833          & 0.039 & I  & I       & BLS1 &   &   \\
MARK871          & 0.034 & S  & Sb      & BLS1 & 3690 & M03 \\
MARK885          & 0.026 & SB & SBb     & BLS1 &   &   \\
MARK896          & 0.027 & S  & Sc      & NLS1 & 1135 & V01 \\
MARK915          & 0.025 & S  & Sa      & BLS1 &   &   \\
MARK1040         & 0.016 & S  & Sb      & NLS1 & 1830 & O82 \\
MARK1044         & 0.016 & S  & Sa      & NLS1 & 1010 & V01 \\
MARK1126         & 0.010 & S  & Sb      & BLS1 &   &   \\
MARK1218         & 0.028 & SB & SBa     & BLS1 &   &   \\
MARK1330         & 0.009 & S  & Sb/c    & BLS1 &   &   \\
MARK1376         & 0.007 & ?  & edge on & BLS1 &   &   \\
MARK1400         & 0.029 & S  & Sa      & BLS1 &   &   \\
MARK1469         & 0.031 & S  & Sa      & BLS1 &   &   \\
MS1110+2210      & 0.030 & E  & E       & BLS1 &   &   \\
NGC235           & 0.022 & S  & Sa/b    & BLS1 &   &   \\
NGC526A          & 0.018 & I  & E/S0    & BLS1 &   &   \\
NGC1019          & 0.024 & SB & SBb     & BLS1 &   &   \\
NGC1566          & 0.004 & S  & Sb      & BLS1 & 2580 & K91 \\
NGC2639          & 0.011 & S  & Sb      & BLS1 & 3100 & H97 \\
NGC3227          & 0.003 & ?  & ?       & BLS1 & 3920 & S90 \\
NGC3516          & 0.009 & S  & S0      & BLS1 & 4760 & C86 \\
NGC3783          & 0.009 & SB & I       & BLS1 & 2980 & S90 \\
NGC4051          & 0.002 & SB & Sb      & NLS1 & 1120 & V01 \\
NGC4235          & 0.007 & S  & ?       & BLS1 & 7600 & H97 \\
NGC4748          & 0.014 & S  & Sa      & NLS1 & 1565 & V01 \\
NGC5252          & 0.022 & S  & S0      & BLS1 & 2500 & A96 \\
NGC5548          & 0.017 & S  & Sa      & BLS1 & 5610 & S90 \\
NGC5674          & 0.025 & SB & SBc     & BLS1 &   &   \\
NGC5940          & 0.033 & SB & SBc     & BLS1 & 5240 & M03 \\
NGC6104          & 0.028 & SB & SBb     & BLS1 &   &   \\
NGC6212          & 0.030 & S  & Sb      & BLS1 & 6050 & H86 \\
NGC6860          & 0.015 & S  & Sb      & BLS1 & 3900 & W92 \\
NGC7213          & 0.006 & S  & Sa      & BLS1 & 3200 & W92 \\
NGC7314          & 0.006 & S  & Sd      & BLS1 &   &   \\
NGC7469          & 0.017 & S  & Sb/c    & BLS1 & 3460 & C86 \\
IIZW10           & 0.034 & S  & ?       & BLS1 & 3760 & M03 \\
PKS0518-458      & 0.034 & E  & E       & BLS1 & 5020 & M03 \\
TOL1059+105      & 0.034 & S  & S0      & BLS1 &   &   \\
TOL2327-027      & 0.033 & S  & SB      & BLS1 &   &   \\
UM146            & 0.017 & S  & SBa     & BLS1 &   &   \\
UGC3223          & 0.018 & S  & S(B)b/c & BLS1 & 4740 & S90 \\
WAS45            & 0.024 & SB & Sa      & BLS1 &   &   \\
UGC10683B        & 0.031 & SB & SBa     & BLS1 &   &   \\
UGC12138         & 0.025 & SB & SBa     & BLS1 &   &   \\
UM614            & 0.033 & S  & S0      & BLS1 &   &   \\
X0459+034        & 0.016 & E  & E       & BLS1 & 4320 & G82 \\
IZW1$^b$             & 0.061 & SB & -       & NLS1 & 1400 & O77 \\
MS0144.2-0055$^b$    & 0.080 & S  & -       & NLS1 & 1100 & V01 \\
MARK705$^c$          & 0.028 & SB & -       & NLS1 & 1790 & V01 \\
MS1217.0+0700$^b$    & 0.080 & S  & -       & NLS1 & 1765 & V01 \\
MS1519.8-0633$^b$    & 0.084 & SB & -       & NLS1 & 1115 & V01 \\
MS2210.2+1827$^b$    & 0.079 & SB & -       & NLS1 &  690 & V01 \\
\tablenotetext{a}{References for FWHM --
A96: Acosta-Pulido, et al. 1996,
C86: Crenshaw 1986,
G82: Ghigo, et al. 1982,
G91: Gregory, Tifft, \& Cocke 1991,
H86: Halpern \& Filippenko 1986,
H97  Ho et al. 1997,
K83: Kollatschny \& Fricke 1983,
K00: Kollatschny, Bischoff, \& Dietrich 2000,
K91: Kriss et al. 1991,
M03: Marziani et al. 2003,
O77: Osterbrock 1977,
O81: Osterbrock 1981,
O82: Osterbrock \& Shuder 1982,
R00: Rodriguez-Ardila, Pastoriza, \& Donzelli 2000,
S90: Stirpe 1990,
V01: V\'eron-Cetty, V\'eron, \& Gon\c{c}alves 2001,
W92: Winkler 1992
}
\tablenotetext{b}{Observed with the F814W filter.}
\tablenotetext{c}{Observed with the F547M filter.}
\enddata
\end{deluxetable}
\normalsize

\clearpage
\begin{figure}
\plotone{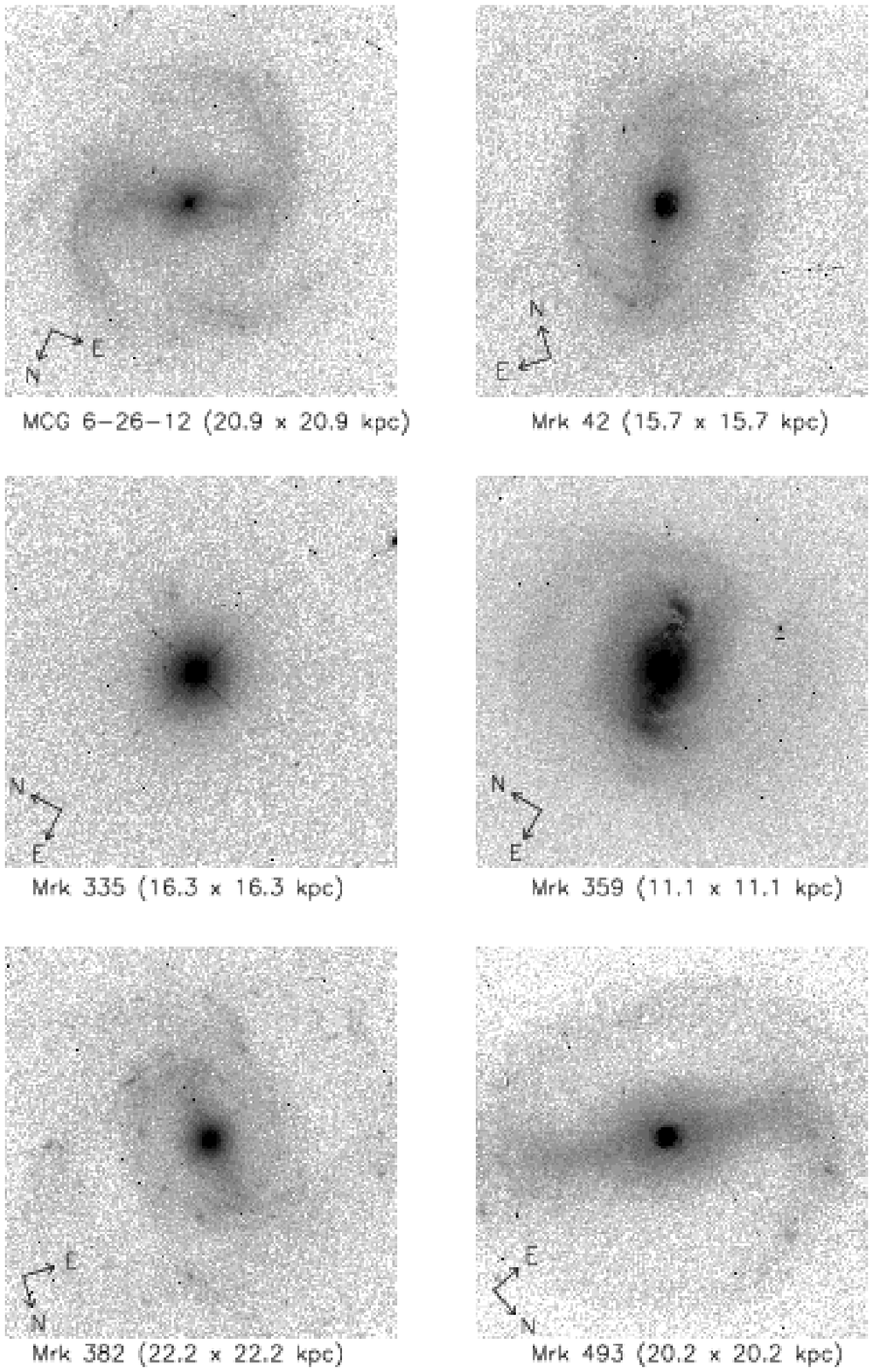}
\\Fig.~1.
\end{figure}

\clearpage
\begin{figure}
\plotone{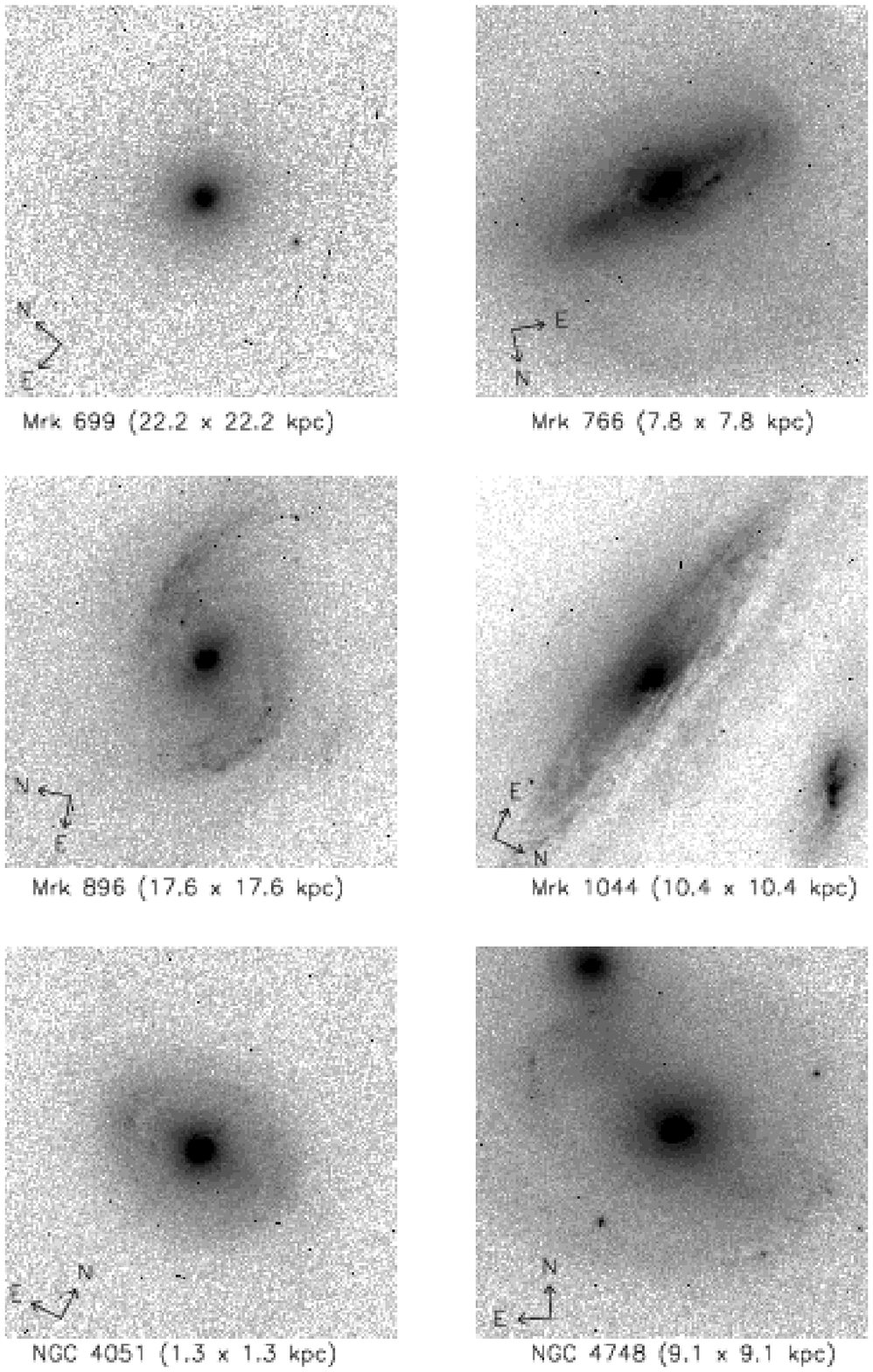}
\\Fig.~1 -- Continued.
\end{figure}

\clearpage
\begin{figure}
\plotone{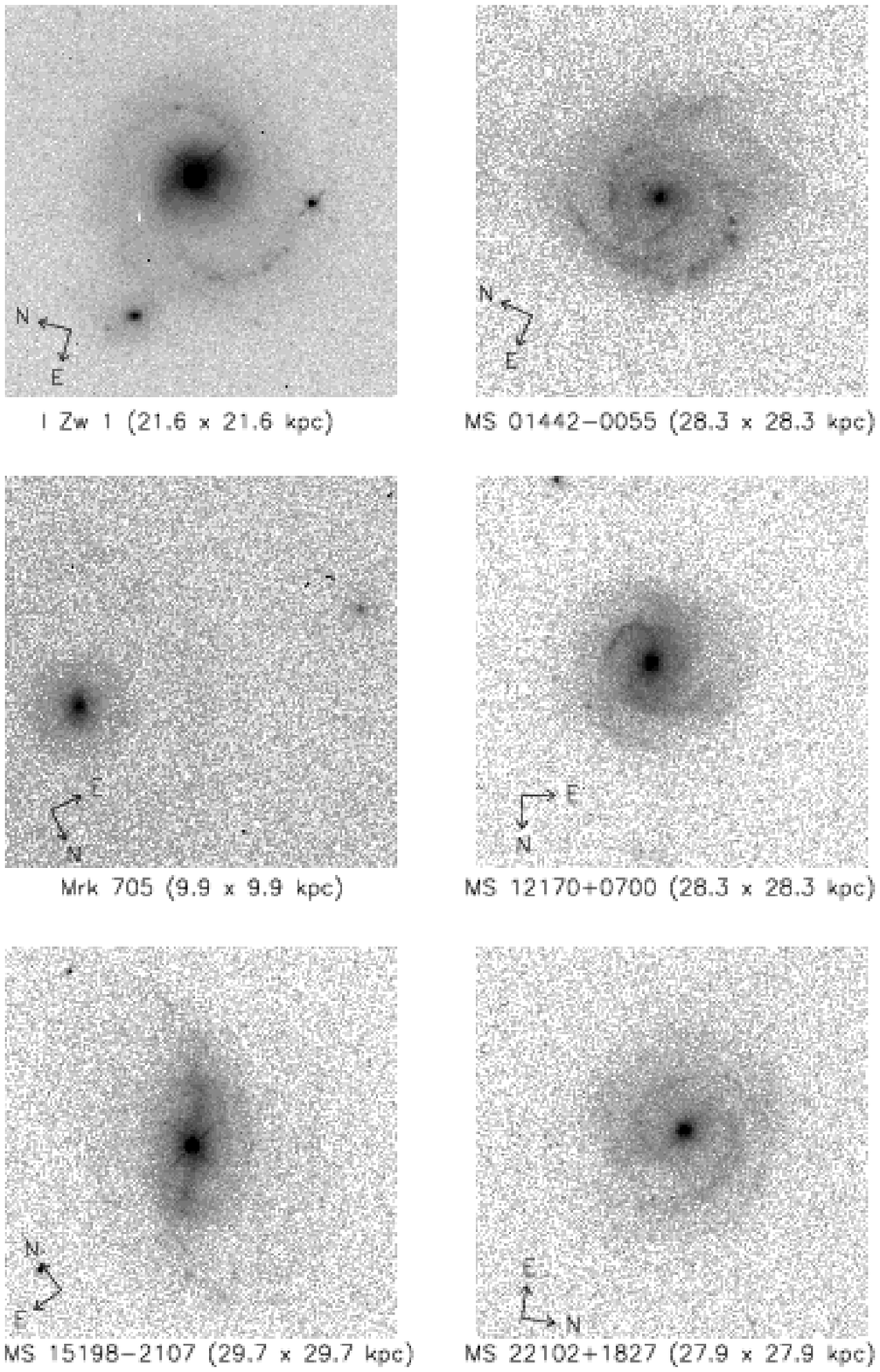}
\\Fig.~2.
\end{figure}

\clearpage
\begin{figure}
\rotatebox{90}{
\epsscale{0.8}
\plotone{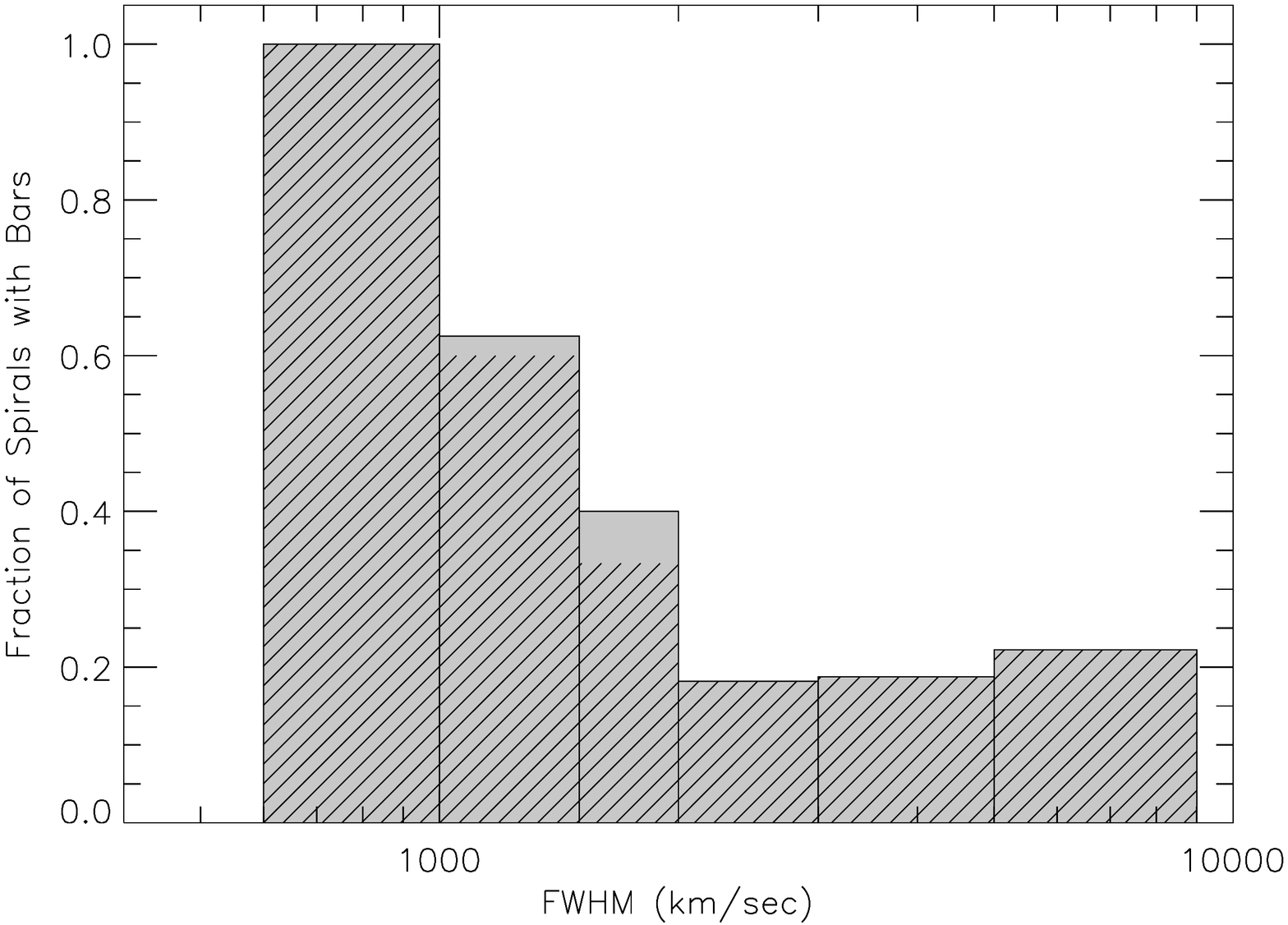}}
\\Fig.~3.
\end{figure}

\end{document}